# XANES-Guided Molecular Structure Generation via Conditional Diffusion and Forward Spectral Prediction

*Yu Fujikata *[ab] and Teruyasu Mizoguchi *[a]*

[a] Institute of Industrial Science, The University of Tokyo, Tokyo, 153-8505, Japan.

[b] Mitsubishi Chemical Corporation, Science & Innovation Center, Yokohama, 227-8502, Japan.

**Abstract**

Reconstructing complete molecular structures from X-ray absorption near-edge structure (XANES) spectra remains challenging because the spectra primarily encode local chemical environments, with overlapping contributions from multiple absorbing sites. Here, we develop an integrated inverse–forward framework combining XANES-conditioned diffusion-based molecular structure generation with forward spectral prediction for candidate selection. The inverse model generates atomic species, bond connectivity, and three-dimensional coordinates from C, N, and O K-edge XANES, given only the total number of atoms and without molecular formulas or element-specific atom counts. Integrating all three absorption edges substantially improves structural recovery, achieving a best-of-10 exact-match rate of 56.6% across 2,888 test molecules. Beyond exact matches, the generated candidates reproduce the functional groups and structural motifs present in the reference molecules, with a macro-averaged F1 score of 0.951 across 21 categories. To select among the generated candidates, we rank them by the agreement between input and forward-predicted spectra increases the exact-match rate from 24.3% for unranked top-1 selection to 43.9%. The same spectral-consistency criterion also relaxes the remaining requirement of a known atom count: comparing candidates generated under seven different total atom counts identifies the correct count for 65.6% of molecules. Application to experimental core-excitation spectra yields an exact structural match for imidazole and recovers characteristic functional groups in two additional molecules, while revealing limitations associated with simulation–experiment spectral discrepancies. These findings establish an inverse–forward approach for translating complementary element-specific spectroscopic information into whole-molecule structure generation and selection, providing a foundation for molecular structure elucidation from experimental XANES.

## 1. Introduction

Elucidating the chemical structures and three-dimensional (3D) atomic arrangements of organic molecules is essential for understanding the properties and functions of molecular materials. Conventional characterization techniques, including nuclear magnetic resonance (NMR) spectroscopy, mass spectrometry (MS), and infrared (IR) spectroscopy, provide complementary information on molecular connectivity, mass and fragmentation patterns, and vibrational signatures. For thin films, adsorbed layers, interfaces, and surface-functionalized materials, however, element-specific characterization of local electronic and bonding environments is particularly valuable. X-ray absorption near-edge structure (XANES) spectroscopy provides such information by probing electronic transitions from core levels to unoccupied states[1]. In particular, soft X-ray XANES at the C, N, and O K-edges, commonly referred to as near-edge X-ray absorption fine structure (NEXAFS) spectroscopy, is sensitive to the chemical bonding and functional groups of organic molecules[2,3]. Its applications include surface-sensitive measurements using electron-yield detection and molecular orientation analysis using linearly polarized X-rays[4,5]. Moreover, XANES is applicable to amorphous and heterogeneous materials and, with suitable experimental configurations, enables *in situ* and *operando* characterization of element-specific chemical environments[6,7]. These capabilities make XANES a valuable complement to conventional molecular characterization techniques.

Despite these advantages, extracting complete molecular structures from XANES spectra remains challenging[8,9]. Whereas complementary information from NMR, MS, and IR spectroscopy can be combined to elucidate molecular connectivity and structural features, XANES primarily probes the local electronic and bonding environments of absorbing atoms. Conventional XANES analysis therefore relies largely on reference spectra, spectral feature assignments, and first-principles calculations to identify local chemical environments[10,11]. However, distinct molecules with similar local bonding environments may exhibit similar spectral features, making the spectrum-to-structure relationship potentially nonunique[12]. In addition, whole-molecule XANES spectra contain superimposed contributions from multiple absorption sites, complicating the assignment of spectral features to individual atomic environments[13]. Exhaustive comparison of candidate structures using first-principles XANES calculations is also computationally demanding[14–17]. An efficient approach for directly generating and evaluating molecular structures from XANES spectra is therefore highly desirable.

Recent advances in machine learning have begun to address these limitations by enabling direct reconstruction of atomic structures from spectroscopic information[18]. Kwon et al. demonstrated XANES-guided diffusion-based generation of three-dimensional atomic structures of amorphous carbon[19]. Okubo et al. developed an equivariant diffusion model conditioned on electron energy-loss near-edge structure (ELNES)/XANES spectra to generate local 3D atomic coordination environments[20]. Zhong et al. similarly proposed a framework that aligns X-ray absorption spectroscopy (XAS) and local structural representations to generate 3D atomic coordinates given known atomic species and atom counts[21]. These studies demonstrate that core-excitation spectra can constrain explicit atomic arrangements beyond conventional structural descriptors, such as coordination numbers and bond lengths. Nevertheless, extending this capability to entire molecules requires the simultaneous determination of atomic species, bond connectivity, and 3D atomic coordinates, rather than only the local arrangement around a specified absorbing atom. A central challenge is therefore to integrate the local chemical information encoded in XANES spectra into a consistent molecular connectivity and 3D structure, particularly when contributions from multiple absorption sites are superimposed.

Spectroscopy-conditioned generative models provide a promising route toward this goal. Wang et al. introduced DiffSpectra, which combines a Transformer-based spectral encoder (SpecFormer) with a Diffusion Molecule Transformer (DMT) to generate molecular structures, including connectivity and 3D atomic arrangements, from multiple spectroscopic modalities[22]. By sampling multiple structural candidates from a given spectrum, diffusion models offer a natural way to address the nonuniqueness of spectral inverse problems[23]. For XANES, additional structural constraints can be obtained by combining multiple absorption edges, as the C, N, and O K-edges provide complementary information on the bonding environments of different elements[24]. Integrating these element-specific signatures may therefore improve the reconstruction of complete molecular structures beyond what is achievable using a single absorption edge[25].

A further challenge is identifying the most plausible structure among the generated candidates. This can be addressed by coupling inverse structure generation with forward spectral prediction[26]. Specifically, a diffusion model generates candidate molecular structures conditioned on the observed XANES spectra, while a forward machine-learning model predicts the corresponding spectra and ranks the candidates according to their spectral agreement. This inverse–forward strategy offers a means of selecting structures from an otherwise ambiguous candidate space.

Here, we develop an integrated inverse–forward framework for whole-molecule structure prediction from XANES spectra (Figure 1). In the inverse step, a conditional diffusion model generates atomic species, bond connectivity, and 3D coordinates from C, N, and O K-edge XANES spectra, given only the total number of atoms and without molecular formulas or element-specific atom counts. We assess the contribution of each absorption edge and show that combining their complementary information substantially improves structural recovery. In the forward step, a machine-learning model predicts the XANES spectra of the generated candidates, and the candidates are ranked by their agreement with the input spectra. We further show that the same spectral-consistency criterion can identify the total atom count, relaxing the last compositional input required for generation, and finally apply the framework to experimental core-excitation spectra. By coupling diffusion-based inverse generation with forward spectral evaluation, this framework enables both the reconstruction and the selection of whole-molecule structures from element-specific spectroscopic information.

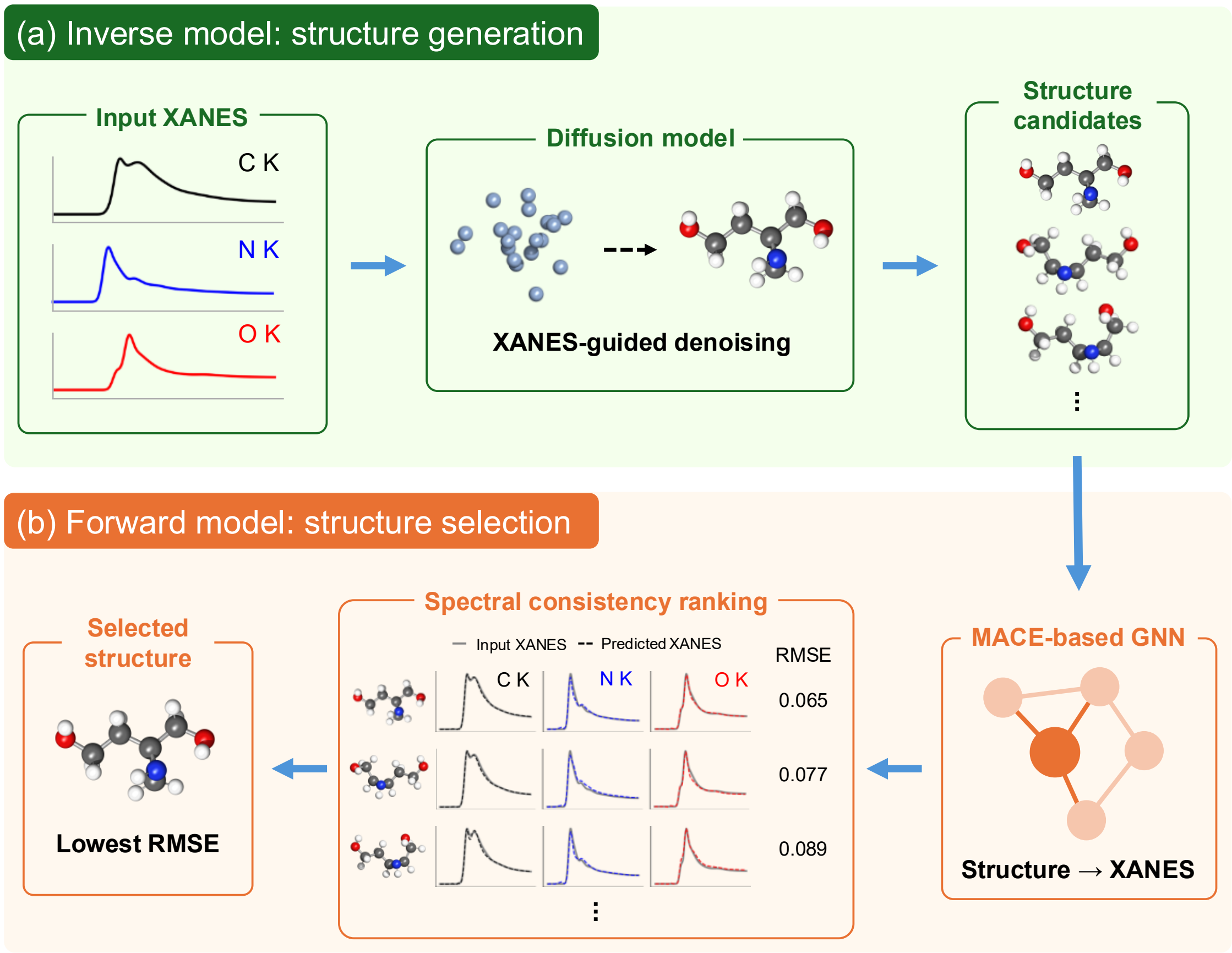

**Figure 1.** Overview of the inverse–forward framework for XANES-guided molecular structure generation and selection. (a) Inverse model: A conditional diffusion model generates multiple three-dimensional molecular structure candidates from input C, N, and O K-edge XANES spectra through an iterative denoising process. (b) Forward model: A MACE-based graph neural network predicts the XANES spectra of each generated candidate from its three-dimensional atomic structure. The candidates are ranked according to the root-mean-square error (RMSE) between the predicted and input spectra, averaged over the available absorption edges, and the candidate with the lowest RMSE is selected as the most plausible molecular structure.

## 2. Methods

### 2.1. Molecular Dataset and Data Splitting

A dataset of 27,112 organic molecules derived from QM9[27,28] was used for molecular structure generation and XANES calculations. The molecules contained H, C, N, O, and F atoms, with total atom counts ranging from 4 to 26, including hydrogen atoms, and encompassed a diverse range of functional groups and structural motifs (Figure S1). The dataset was randomly split at the molecular level into training (16,267 molecules), validation (5,422 molecules), and test (5,423 molecules) sets. To ensure a consistent comparison of the inverse model across different absorption-edge conditions, a common subset of 2,888 test molecules containing C, N, and O atoms was used for performance evaluation.

### 2.2. First-Principles XANES Calculations and Spectral Preprocessing

The C, N, and O K-edge XANES spectra were calculated using density functional theory (DFT) as implemented in CASTEP[29]. The Perdew–Burke–Ernzerhof (PBE) exchange–correlation functional[30] and ultrasoft pseudopotentials[31] were employed, consistent with the computational settings of a previously reported molecular K-edge spectral database[32]. The core-hole pseudopotential approach has been validated for K-edge spectra of solids and gas-phase molecules[33], while the plane-wave basis provides a systematic representation of unoccupied electronic states. Each molecule was placed in a cubic supercell with a side length of 15 Å, using a plane-wave cutoff energy of 500 eV and Γ-point-only sampling. Core excitations were described using the excited core-hole (XCH) approximation, in which a 1s core hole was introduced at the absorbing atom and the excited electron was placed in the lowest conduction state to maintain charge neutrality[34]. Transition probabilities were calculated using the ELNES module of CASTEP, and the resulting spectra were broadened with a Gaussian function with a standard deviation of 0.5 eV.

Transition energies were evaluated using the total-energy difference between the core-excited and ground states, together with an atomic core-level correction, following the previously established procedure[33]: An additional constant energy offset was applied to each absorption edge based on comparisons with experimental spectra[34].

The calculated spectra were interpolated onto 300-point energy grids with a spacing of 0.1 eV, covering 280.0–309.9 eV for the C K-edge, 395.0–424.9 eV for the N K-edge, and 525.0–554.9 eV for the O K-edge. Whole-molecule XANES spectra were obtained by summing the contributions from all absorbing sites of each element, followed by normalization to unit maximum intensity. For molecules lacking a given element, the corresponding spectral input was set to a 300-point zero vector.

### 2.3. XANES-Conditioned Molecular Structure Generation

Molecular structures were generated using a conditional diffusion model based on the DiffSpectra framework[22]. A Transformer-based spectral encoder, SpecFormer[35], extracted spectral features from 300-point XANES spectra using a patch size of 10 and a stride of 5. A Diffusion Molecule Transformer (DMT) [36,37] was employed to jointly generate atomic species, formal charges, bond information, and three-dimensional (3D) atomic coordinates. The DMT consisted of six Transformer blocks with a hidden dimension of 256, 16 attention heads, and a dropout rate of 0.1. Self-conditioning was incorporated into the diffusion process[38].

Separate models were trained using individual C, N, or O K-edge spectra or their combination. An unconditional model without spectral conditioning was also trained to assess the contribution of XANES information. For all generation conditions, the total number of atoms, including hydrogen, was specified, whereas molecular formulas and element-specific atom counts were not provided.

The diffusion models were trained to predict the original molecular structure, $x_0$, from its noisy representation. The training objective comprised reconstruction losses for atomic coordinates, atomic species and formal charges, and bond information, with respective weights of 1.00, 0.25, and 0.10. A cosine noise schedule[39] and the AdamW optimizer[40] were employed. The learning rate was linearly increased to $2 \times 10^{-4}$ over the first 100,000 optimizer steps and held constant thereafter. Training was performed with a batch size of 512, and an exponential moving average (EMA) of model parameters was maintained with a decay factor of 0.999. Model checkpoints were selected based on validation performance.

During inference, 1,000-step ancestral sampling was performed to generate ten molecular structure candidates per input condition using different random noise realizations.

### 2.4. Evaluation of Generated Molecular Structures and Functional Groups

Molecular connectivity was reconstructed from the generated atomic species and 3D coordinates using Open Babel[41], and the resulting structures were validated using RDKit. Exact molecular matches were determined by comparing the full Standard InChIKeys[42] of the generated and reference molecules. Candidates that failed structure reconstruction or validation were counted as incorrect.

Structural similarity was evaluated using Morgan Tanimoto[43], maximum common substructure (MCS)[44], and Fraggle similarities. Morgan fingerprints were calculated with a radius of 2 and 2,048 bits. The MCS was identified by matching atomic species, bond orders, valences, and ring structures. The MCS similarity was defined using a Jaccard-type coefficient[45]:

$$S_{\mathrm{MCS}} = \frac{N_{\mathrm{MCS}}}{N_{\mathrm{ref}} + N_{\mathrm{gen}} - N_{\mathrm{MCS}}}$$

where $N_{\mathrm{ref}}$ and $N_{\mathrm{gen}}$ are the numbers of non-hydrogen atoms in the reference and generated molecules, respectively, and $N_{\mathrm{MCS}}$ is the number of atoms in their maximum common substructure.

Functional groups and structural motifs were identified using SMARTS patterns[46] implemented in RDKit. For each reference molecule, the union of functional groups identified across the ten generated candidates was used as the predicted label set. Multilabel classification performance was evaluated for 21 categories, and the macro-F1 score was calculated as the arithmetic mean of the category-wise F1 scores.

### 2.5. Forward XANES Prediction

The forward model employed a graph neural network based on the MACE architecture[47]. Because XANES features depend sensitively on the three-dimensional local geometry around each absorbing atom, including bond angles and conjugation, we adopted MACE, which constructs E(3)-equivariant many-body representations of local atomic environments and has demonstrated high accuracy and data efficiency in learning structure–property relationships. This architecture also allows the generated 3D candidates to be evaluated directly from their atomic coordinates, while yielding rotationally invariant spectral predictions. Atomic numbers and three-dimensional Cartesian coordinates were used as inputs to construct a neighbor graph with a cutoff radius of 6.0 Å. Interatomic distances were expanded using 32 Gaussian radial basis functions, while angular information was represented by spherical harmonics. The network consisted of four MACE blocks with 96 feature channels. Atomic features from each block were mean-pooled over the molecule, concatenated, and passed through a regression network to predict a 300-point whole-molecule XANES spectrum. Separate models were trained for the C, N, and O K-edges.

The forward models used the same molecule-level data split as the inverse models, restricted to molecules containing the corresponding absorbing element. Training minimized the mean squared error (MSE) between predicted and reference spectral intensities using the AdamW optimizer[40].

For candidate ranking, the XANES spectra of generated molecular structures were predicted using the forward models and compared with the input spectra using the root-mean-square error (RMSE). When multiple absorption edges were available, the RMSE values were averaged with equal weights, and candidates were ranked in ascending order of the resulting RMSE. For atom-count selection, the mean RMSE across ten generated candidates was calculated for each atom-count condition, and the condition with the lowest mean RMSE was selected.

## 3. Results and Discussion

### 3.1. Inverse Analysis: Molecular Structure Generation from XANES

We first evaluated the ability of the inverse model to generate whole-molecule structures from XANES spectra, focusing on the contributions of element-specific spectral information. Structure generation was compared using individual C, N, and O K-edge spectra and their combination (Figure 2). An unconditional model, which generated molecules using only the total atom count, was included to assess the contribution of spectral conditioning. We also evaluated local structure generation from site-specific XANES, obtaining results consistent with previous studies[20,21] (Supporting Information).

The ability to recover the reference molecular structures was quantified using the best-of-*k* exact-match rate based on Standard InChIKey, evaluated on a common set of 2,888 test molecules (Figure 2a). With ten generated candidates, the C, N, and O K-edge models achieved exact-match rates of 36.6%, 27.6%, and 33.6%, respectively, compared with only 0.2% for unconditional generation. Notably, these results were obtained without specifying molecular formulas or element-specific atom counts, despite the diversity of molecular compositions and sizes in the dataset. The substantial improvement over unconditional generation demonstrates that even a single absorption edge provides sufficient chemical information to constrain whole-molecule connectivity, despite the superposition of contributions from multiple absorbing sites.

Combining the C, N, and O K-edges further improved the exact-match rate to 24.3%, 49.0%, and 56.6% for one, five, and ten candidates, respectively, consistently outperforming the single-edge models (Figure 2a). Thus, the combined model recovered the correct molecular identity for more than half of the test molecules within ten candidates, using only the total atom count and XANES spectra as structural constraints. This improvement highlights the complementary information provided by different absorption edges: the C K-edge probes carbon bonding environments, whereas the N and O K-edges provide additional constraints on the corresponding heteroatom environments. Integrating these element-specific signatures therefore reduces structural ambiguity and enables more accurate reconstruction of molecular connectivity.

To assess structural recovery beyond exact molecular identity, we compared generated and reference molecules using Morgan Tanimoto, maximum common substructure (MCS), and Fraggle similarities (Figure 2b). The combined C, N, and O K-edge model achieved mean best-of-10 similarities of 0.717, 0.902, and 0.688, respectively. The corresponding values were 0.571, 0.849, and 0.618 for the C K-edge; 0.499, 0.820, and 0.572 for the N K-edge; 0.548, 0.842, and 0.602 for the O K-edge; and 0.173, 0.634, and 0.366 for unconditional generation. The combined model consistently achieved the highest similarity across all three metrics, indicating that multi-edge conditioning improves not only exact molecular recovery but also the preservation of structural features. In particular, the high MCS similarity suggests that generated molecules frequently retain substantial portions of the reference molecular scaffold, even when an exact match is not obtained.

We further evaluated the recovery of 21 functional groups and structural motifs (Figure 2c). The combined C, N, and O K-edge model achieved a macro-F1 score of 0.9509 (95% bootstrap confidence interval: 0.9420–0.9592), substantially exceeding those of the individual C (0.8091), N (0.7220), and O (0.7914) K-edge models and unconditional generation (0.2678). Notably, the combined model outperformed all single-edge models across all 21 categories, with particularly pronounced improvements for esters (0.908), epoxides (0.947), and lactones (0.839). These results suggest that complementary information from different absorption edges provides stronger constraints on the chemical environments and connectivity of functional groups than any individual edge alone. Thus, multi-edge conditioning enables more accurate recovery of diverse chemical functionalities, even when the complete molecular structure is not reproduced.

Overall, these results demonstrate that whole-molecule XANES spectra contain sufficient complementary chemical information to guide the generation of chemically consistent molecular structures, even when contributions from multiple absorbing sites overlap. Nevertheless, the probabilistic generation process produces multiple candidates with potentially different connectivities, and an exact match was obtained within ten candidates for 56.6% of the test molecules. Identifying the most plausible structure among these candidates therefore remains essential. In the following section, we address this challenge by evaluating spectral consistency-based candidate ranking using a forward XANES prediction model.

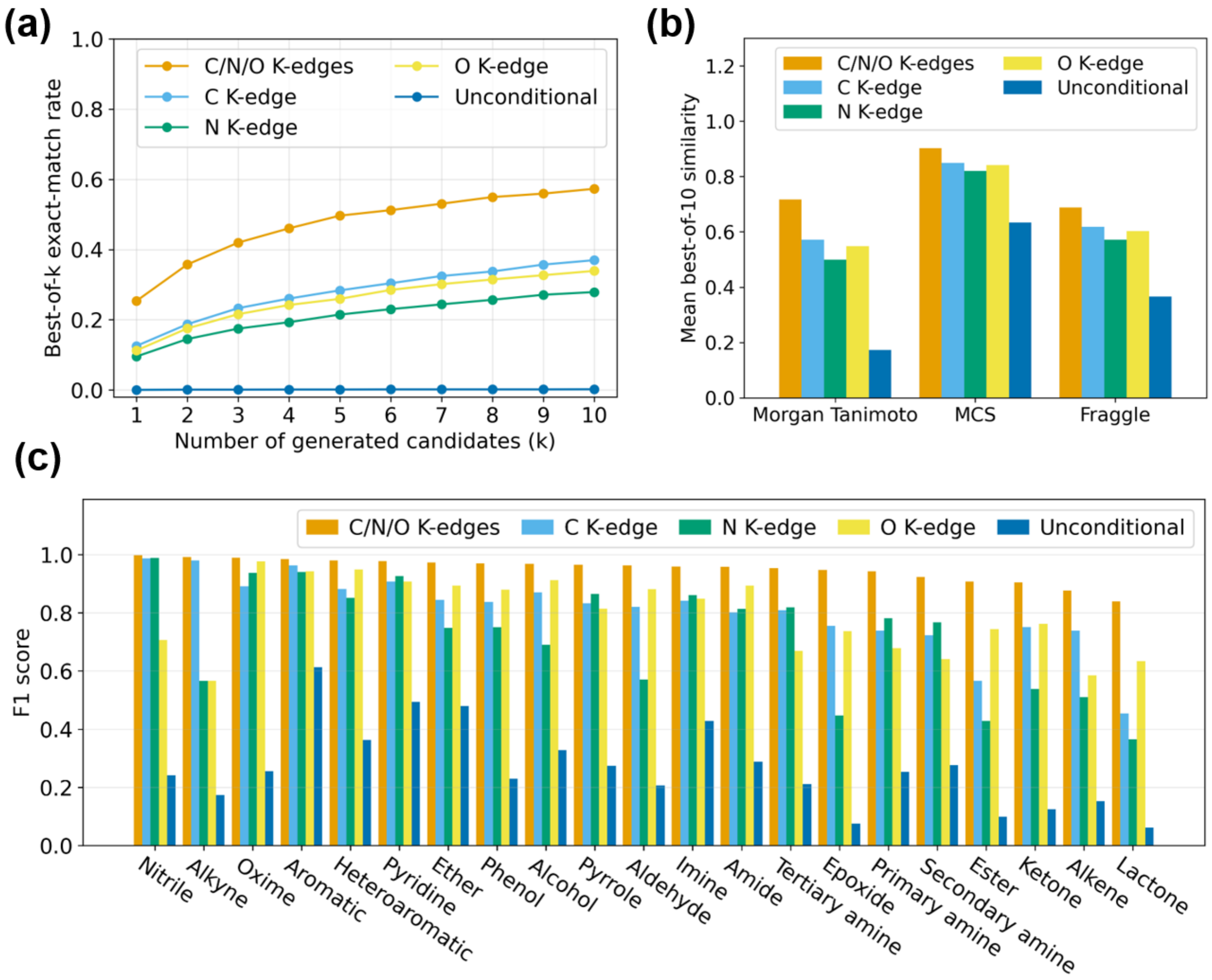

**Figure 2.** Molecular structure generation from single-edge and multi-edge XANES. (a) Best-of-*k* exact-match rates based on Standard InChIKey as a function of the number of generated candidates (*k*). (b) Mean best-of-10 molecular structural similarities evaluated using Morgan Tanimoto, maximum common substructure (MCS), and Fraggle similarity metrics. (c) F1 scores for C-, N-, and O-containing functional groups and structural motifs. Five generation conditions were compared: combined C, N, and O K-edges, C K-edge only, N K-edge only, O K-edge only, and unconditional generation.

## 3.2. Forward Analysis: Spectral Consistency-Based Ranking of Generated Candidates

Although the inverse model successfully generated reference-matching structures among multiple candidates, the best-of-*k* evaluation does not indicate which candidate should be selected when the reference structure is unknown. To address this limitation, we evaluated a spectral consistency-based ranking approach using a forward model that predicts XANES spectra from generated molecular structures (Figure 1). The MACE-based forward model achieved mean RMSE values of 0.0459, 0.0437, and 0.0413 for the C, N, and O K-edges, respectively, with corresponding median values of 0.0418, 0.0402, and 0.0374 on the test sets (Figures S5–S7). The predicted spectra reproduced the principal absorption features and overall spectral profiles, enabling efficient evaluation of generated candidates without performing additional first-principles calculations.

To determine whether spectral agreement reflects structural similarity, we examined the relationship between the XANES reconstruction RMSE and maximum common substructure (MCS) similarity for the generated candidates (Figure 3a). A clear negative correlation was observed, with Pearson and Spearman correlation coefficients of −0.600 and −0.643, respectively. The mean MCS similarity calculated within 30 RMSE quantile bins also decreased with increasing RMSE, indicating that candidates with better spectral agreement generally retained more of the reference molecular structure. Consistent relationships were observed using alternative structural similarity metrics (Morgan Tanimoto and Fraggle) and spectral comparison metrics (Wasserstein distance and cosine similarity; Figure S8). These findings are consistent with the sensitivity of XANES to local electronic and bonding environments and suggest that spectral agreement across multiple absorption edges provides a useful measure of molecular structural similarity.

We next examined whether this relationship could be exploited to rank candidates generated from the same input spectra. The ten candidates for each molecule were ranked in ascending order of their reconstruction RMSE, and the exact-match rate was evaluated at each rank (Figure 3b). The top-ranked candidates achieved an exact-match rate of 43.9%, which decreased nearly monotonically with rank to 8.6% for the tenth-ranked candidates. This trend demonstrates that spectral consistency can effectively prioritize structurally correct candidates without requiring knowledge of the reference molecular structure.

To quantify the benefit of spectral ranking, we compared three selection strategies: unranked top-1 selection, RMSE-based ranking, and oracle best-of-10 selection (Figure 3c). Unranked top-1 selection yielded an exact-match rate of 24.3%, whereas RMSE-based selection increased this rate to 43.9%, corresponding to an improvement of 19.6 percentage points. The oracle best-of-10 rate was 56.6%, indicating that RMSE-based ranking successfully selected the correct structure in approximately 77.6% of cases where at least one correct candidate was generated. The remaining gap to oracle performance suggests that structurally correct candidates do not always yield the lowest spectral errors, potentially owing to spectral similarities between distinct molecular structures and inaccuracies in forward predictions. Nevertheless, these results demonstrate that combining inverse structure generation with forward spectral prediction substantially improves molecular structure selection.

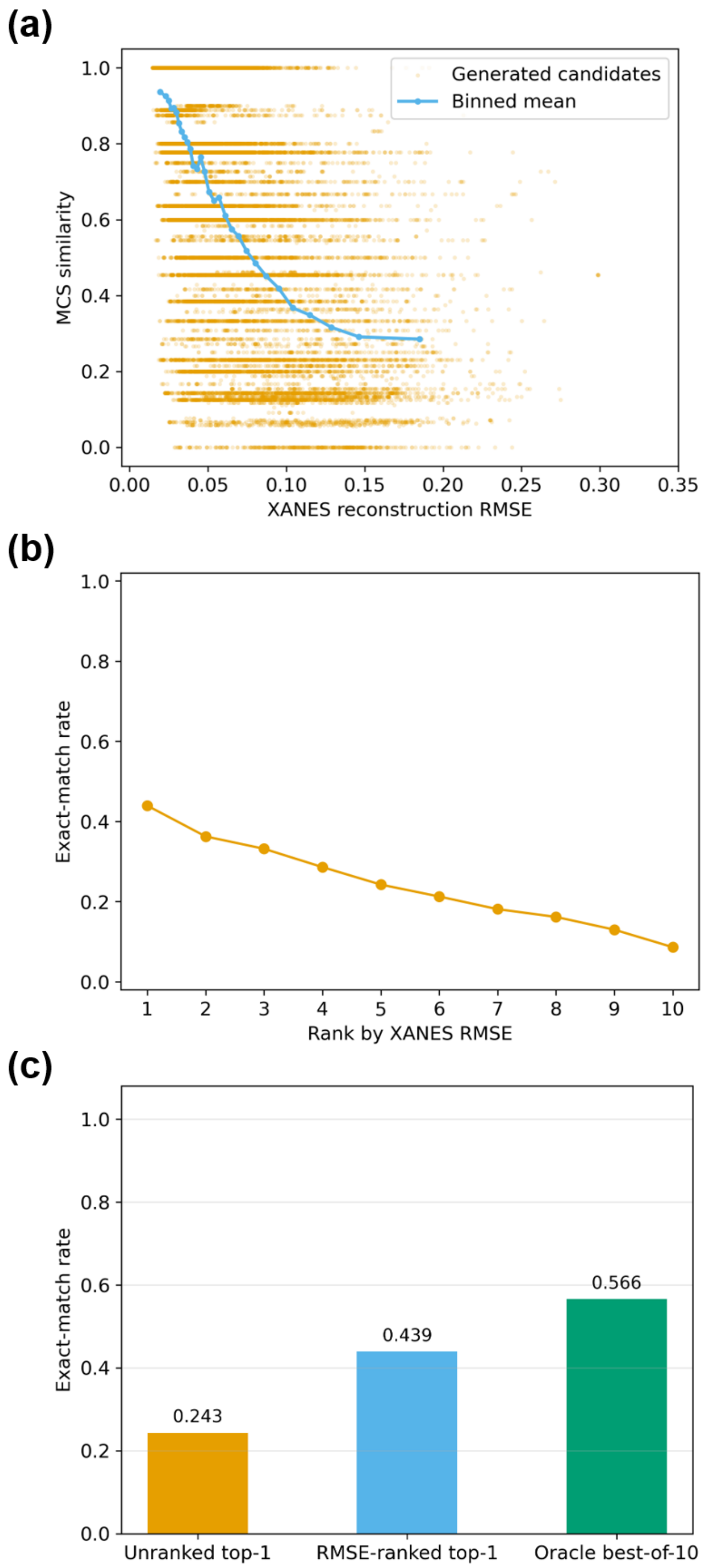

**Figure 3.** Spectral consistency-based ranking and selection of generated molecular structures. (a) Relationship between XANES reconstruction root-mean-square error (RMSE) and maximum common substructure (MCS) similarity to the reference molecule for generated candidates. The line represents the mean MCS similarity within 30 RMSE quantile bins. (b) Exact-match rates at each rank after sorting the ten generated candidates for each input in ascending order of XANES

reconstruction RMSE. (c) Comparison of exact-match rates among unranked top-1, RMSE-ranked top-1, and oracle best-of-10 selection.

### 3.3. Total Atom Count Estimation through Inverse–Forward Analysis

The preceding analyses assumed that the total atom count was known. For unknown molecules, however, this information may not be available. We therefore extended the spectral consistency-based approach to candidates generated with different total atom counts to investigate whether the same criterion could be used to identify the appropriate molecular size. Specifically, the total atom count, including hydrogen atoms, was varied from −3 to +3 relative to the reference molecule, and ten candidates were generated for each of the seven conditions. We evaluated their structural similarities and spectral consistency with the input XANES spectra (Figure 4).

We first examined the maximum common substructure (MCS) similarities of all generated candidates at each atom-count offset (Figure 4a). The highest similarities were generally obtained when the correct atom count was specified, whereas candidates generated with different atom counts still retained substantial portions of the reference molecular scaffold. Interestingly, the ±3 atom conditions exhibited relatively higher similarities than the ±1 and ±2 conditions, indicating a nonmonotonic dependence of structural similarity on atom-count differences. One possible explanation is that the insertion or removal of a methylene ($CH_2$) unit changes the total atom count by three while preserving much of the molecular scaffold and its characteristic functional groups. Such structural modifications could account for the relatively high MCS similarities observed at ±3 atoms. These results suggest that the inverse model can preserve structural features associated with the input XANES spectra even when the specified molecular size differs from that of the reference.

We next investigated whether these structural trends were reflected in spectral consistency. For each atom-count condition, the XANES reconstruction RMSE was averaged over the ten generated candidates (Figure 4b). The correct atom-count condition generally yielded the lowest mean RMSE, consistent with the structural similarity results. Relatively low RMSE values were also observed for the ±3 atom conditions compared with the ±1 and ±2 conditions. Because XANES is primarily sensitive to local electronic and bonding environments, changes in molecular size that preserve characteristic functional groups and their surrounding chemical environments may produce similar spectral features. In particular, the insertion or removal of a $CH_2$ unit may alter the molecular size without substantially affecting the local electronic structures of other functional groups, potentially contributing to the observed nonmonotonic dependence of spectral agreement on atom count[48]. Nevertheless, the overall correspondence between structural similarity and spectral consistency suggests that the reconstruction RMSE provides useful information for distinguishing molecular sizes.

To evaluate atom-count selection quantitatively, we selected the atom-count condition with the lowest mean RMSE among the seven conditions (Figure 4c). The correct total atom count was identified for 3,560 of 5,423 molecules (65.6%), compared with 14.3% expected by random selection among the seven conditions. Among the incorrect selections, ±3 was the most frequent, consistent with the $CH_2$-related ambiguity discussed above. This result demonstrates that spectral consistency can constrain not only the selection of molecular structures but also the total number of atoms, without requiring molecular formulas or element-specific atom counts. Thus, combining inverse structure generation with forward spectral prediction provides a means of extending molecular structure inference beyond a fixed atom-count assumption.

Having established the effectiveness of spectral consistency-based selection across different atom counts, we next applied the inverse–forward framework to experimental core-excitation spectra to assess its applicability to molecular structure prediction without prior knowledge of the total atom count.

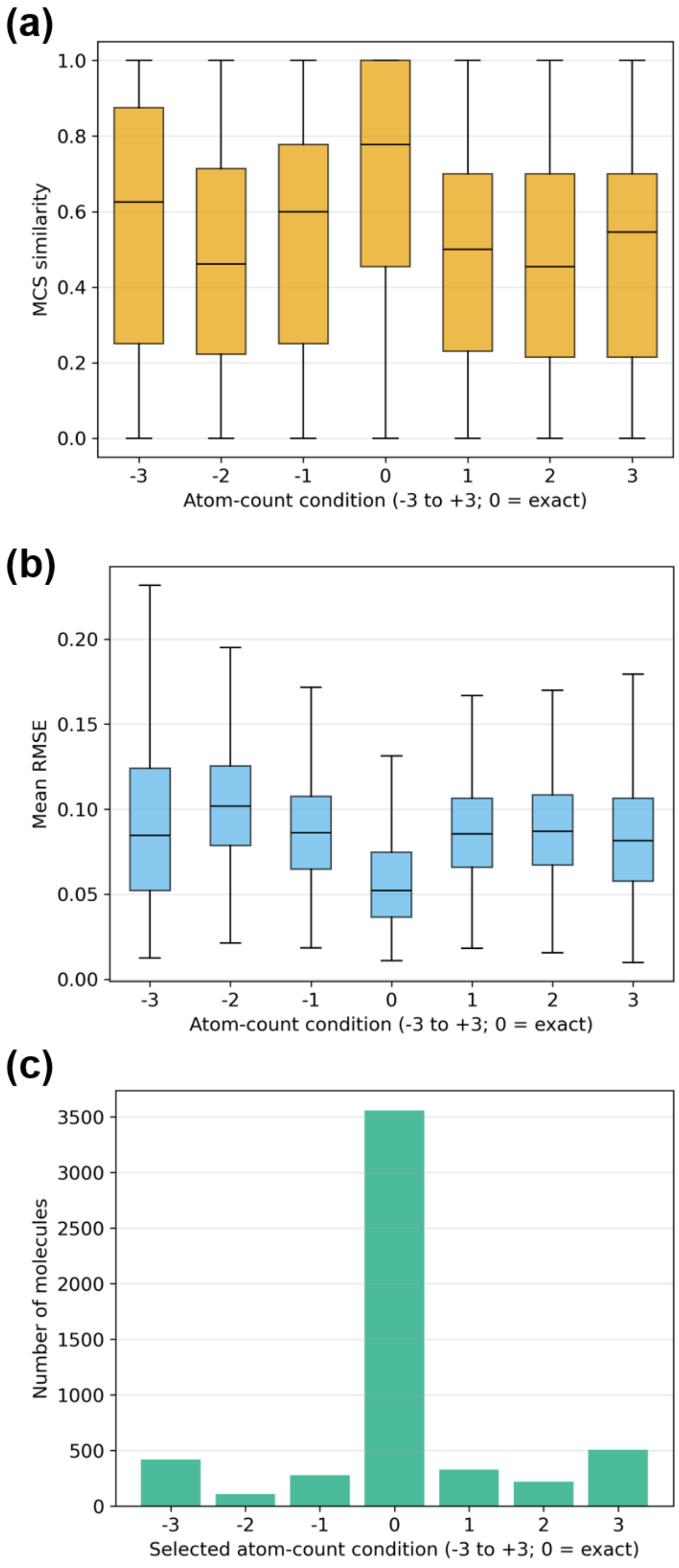

**Figure 4.** Effects of total atom count on molecular structure generation and spectral consistency. The total atom count, including hydrogen atoms, was varied from −3 to +3 relative to the reference molecule ($\Delta N = N_{input} - N_{reference}$). (a) Distributions of maximum common substructure (MCS) similarity for all molecular candidates generated at each atom-count offset. (b) Distributions of XANES reconstruction root-mean-square error (RMSE) averaged over ten generated candidates for each reference molecule at each atom-count offset. (c) Distribution of selected atom-count offsets determined by the lowest mean XANES reconstruction RMSE among the seven conditions.

### 3.4. Molecular Structure Inference from Experimental Core-Excitation Spectra

We finally examined the applicability of the inverse–forward framework, trained on calculated XANES spectra, to experimental core-excitation spectra. Gas-phase electron energy-loss spectroscopy (EELS) spectra of imidazole, acetone, and N-acetylacetamide were obtained from the McMaster University Gas Phase Core Excitation Database[49]. These spectra probe core-electron excitations corresponding to those observed in XANES, providing an opportunity to evaluate the framework using experimental data[50]. For each molecule, the available C, N, and O K-edge spectra were used to generate and evaluate molecular candidates without specifying molecular formulas or element-specific atom counts. Unlike the preceding analysis, in which the atom-count range was defined relative to the reference structure, a common range of 5–20 atoms, including hydrogen, was examined. Ten candidates were generated at each atom count, yielding 160 candidates per molecule, from which the structure with the lowest overall RMSE between the forward-predicted and experimental spectra was selected (Figure 5).

Figure 5a compares the reference structures with the candidates selected from the experimental spectra. For imidazole, the selected structure reproduced the five-membered ring and the connectivity of both nitrogen atoms, yielding an exact match based on Standard InChIKey. For acetone, acetaldehyde was selected instead of the reference structure, although the carbonyl functionality was retained. Similarly, the selected candidate for N-acetylacetamide differed in molecular connectivity but preserved characteristic chemical features, including amide and carbonyl groups. These results demonstrate that the inverse model can generate structures reflecting chemical information contained in experimental core-excitation spectra, despite being trained exclusively on calculated XANES. The exact recovery of imidazole further establishes the feasibility of whole-molecule structure identification from experimental spectra under an unknown atom-count condition.

To assess the spectral consistency of the selected structures, we compared the experimental spectra with forward predictions for both the reference and selected molecular structures (Figure 5b). The predicted spectra of the selected candidates reproduced several characteristic absorption features across the available edges, supporting the use of spectral agreement for candidate evaluation. However, discrepancies remained in relative peak intensities and spectral profiles, particularly at energies farther from the absorption edge. Importantly, such discrepancies were also observed when the correct reference structures were used as inputs to the forward model, indicating that reconstruction errors reflect not only differences in molecular structure but also systematic deviations between calculated and experimental spectra. This limitation was particularly evident for acetone: although the correct molecular structure was present among the generated candidates, RMSE-based ranking favored acetaldehyde. Thus, accurate candidate generation alone does not ensure successful structure identification when spectral inconsistencies affect the ranking process.

The discrepancies between calculated and experimental spectra may arise from several factors, including approximations in electronic-structure and core-excitation calculations, energy alignment, and spectral broadening. Because the forward model was trained on calculated XANES, its predictions may retain systematic deviations from experimental measurements even for structurally correct molecules. To address this issue, Wang et al. reported a machine-learning-based simulation-to-experiment correction approach using 295 pairs of calculated and experimental C, N, and O K-edge spectra of organic molecules[51]. Applying such corrections to forward-predicted spectra could reduce systematic discrepancies and improve the discrimination of candidate structures. Alternatively, adapting the forward and inverse models using experimental spectra may enhance their applicability to measured data. The acetone result particularly highlights the importance of improving spectral consistency between calculation and experiment to translate successful candidate generation into accurate structure selection.

Overall, these results provide a proof of concept for applying the combined inverse–forward framework to experimental core-excitation spectra without prior knowledge of molecular formulas or total atom counts. The exact identification of imidazole and the recovery of characteristic functional groups in the other two molecules demonstrate the potential of experimental spectral information to guide whole-molecule structure generation. At the same time, the limited number of experimental examples and the observed ranking errors underscore the need for further validation and improved agreement between simulated and experimental spectra. With these developments, the proposed framework could provide a foundation for molecular structure elucidation directly from element-specific core-excitation spectroscopy.

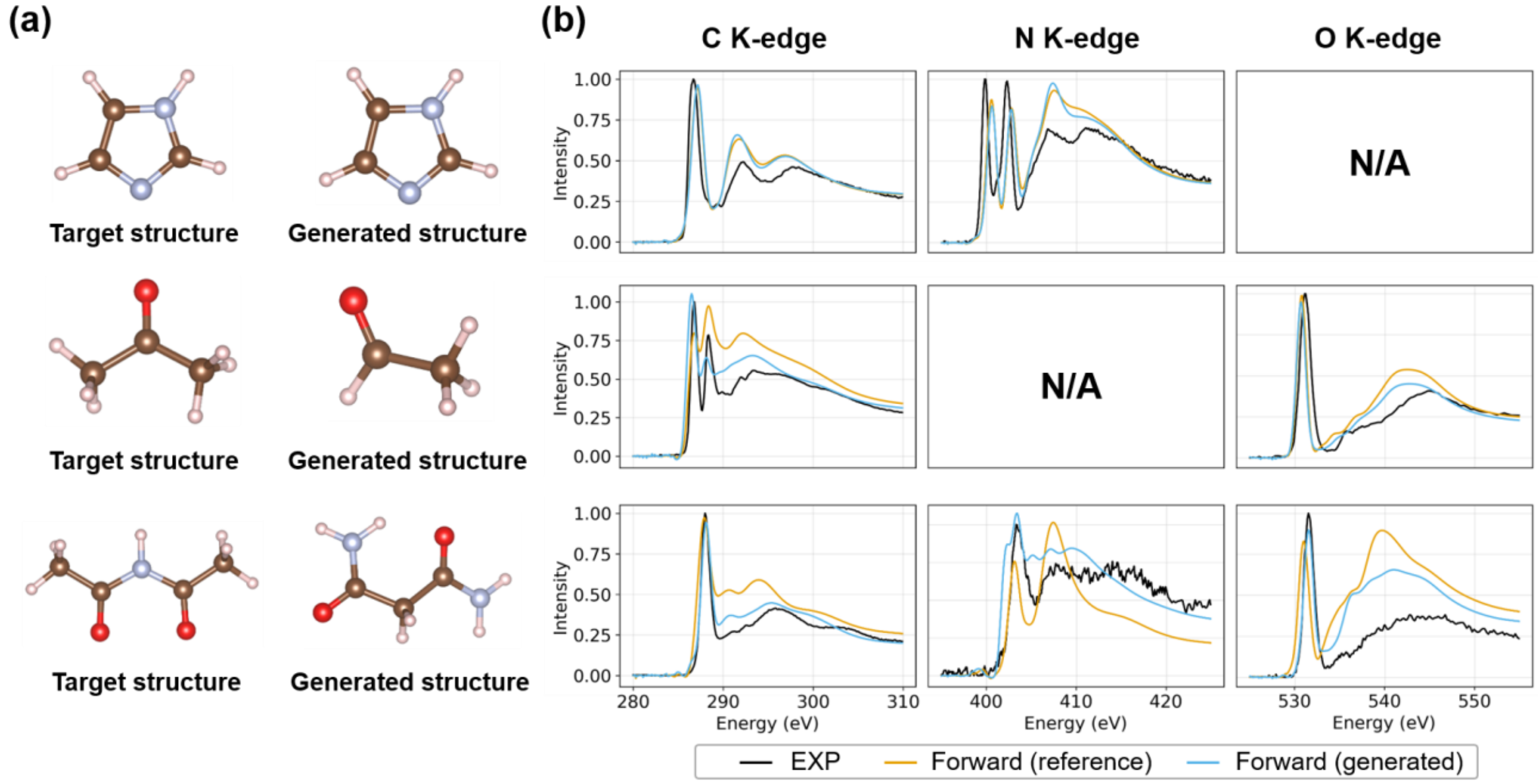

**Figure 5.** Molecular structure generation from experimental core-excitation spectra. (a) Reference molecular structures (left) and selected generated structures (right) for imidazole, acetone, and N-acetylacetamide. H, C, N, and O atoms are shown in light pink, brown, light blue, and red, respectively. (b) Comparison of experimental core-excitation spectra (EXP) with forward-model predictions for the reference structures (Forward (reference)) and selected generated structures (Forward (generated)). The columns correspond to the C, N, and O K-edges from left to right, and the rows correspond to imidazole, acetone, and N-acetylacetamide from top to bottom. N/A indicates that the corresponding element is absent from the molecule.

## 4. Conclusions

In this study, we developed an integrated inverse–forward framework for molecular structure elucidation from XANES spectra, combining conditional diffusion-based structure generation with forward spectral prediction for candidate selection. Given only the total atom count, the inverse model generated atomic species, connectivity, and 3D coordinates from C, N, and O K-edge

XANES, and integrating the three edges achieved a best-of-10 exact-match rate of 56.6% with functional groups largely reproduced even in the absence of an exact match. These results show that whole-molecule XANES, despite the superposition of contributions from multiple absorbing sites, retains sufficient information to constrain molecular connectivity, and that this information is distributed across element-specific edges in a complementary manner.

Forward XANES prediction provided a single criterion, spectral consistency, for decisions that cannot be made from the generative model alone. It increased the top-1 exact-match rate from 24.3% to 43.9% and identified the correct total atom count for 65.6% of molecules, thereby relaxing the last compositional input required for generation. The residual errors are also informative: the frequent confusion of molecules differing by a $CH_2$ unit reflects the intrinsic locality of XANES rather than a deficiency of the generative model, and indicates where complementary information, such as molecular mass, is most effective.

Application to experimental core-excitation spectra yielded an exact match for imidazole and retained characteristic functional groups for the other two molecules. The acetone case, in which the correct structure was generated but not selected, shows that the present bottleneck lies in the agreement between calculated and experimental spectra rather than in candidate generation. Simulation-to-experiment correction, experimental-data adaptation, and extension to additional absorption edges and larger molecular systems are therefore natural next steps. Because XANES is applicable to thin films, interfaces, and in situ environments where conventional techniques are limited, the present framework provides a basis for molecular structure elucidation in settings that have so far relied on local spectral fingerprints alone.

## Supporting Information

Distribution of functional groups and structural motifs in the molecular dataset; methodological details and evaluation results for site-specific XANES-conditioned molecular structure generation; accuracy of forward XANES spectral prediction for the C, N, and O K-edges; and correlations between spectral similarity metrics and molecular structural similarity (PDF).

## Data Availability Statement

The code and datasets used in this study will be made publicly available through GitHub upon publication of this article.

## Notes

The authors declare no competing financial interest.

## Acknowledgment

This study was supported by the Ministry of Education, Culture, Sports, Science and Technology (MEXT).

# Supporting Information

## XANES-Guided Molecular Structure Generation via Conditional Diffusion and Forward Spectral Prediction

**Yu Fujikata*[a,b] and Teruyasu Mizoguchi*[b]**

[a.] **Mitsubishi Chemical Corporation, Science & Innovation Center, Yokohama, 227-8502, Japan.**

[b.] **Institute of Industrial Science, The University of Tokyo, Tokyo, 153-8505, Japan.**

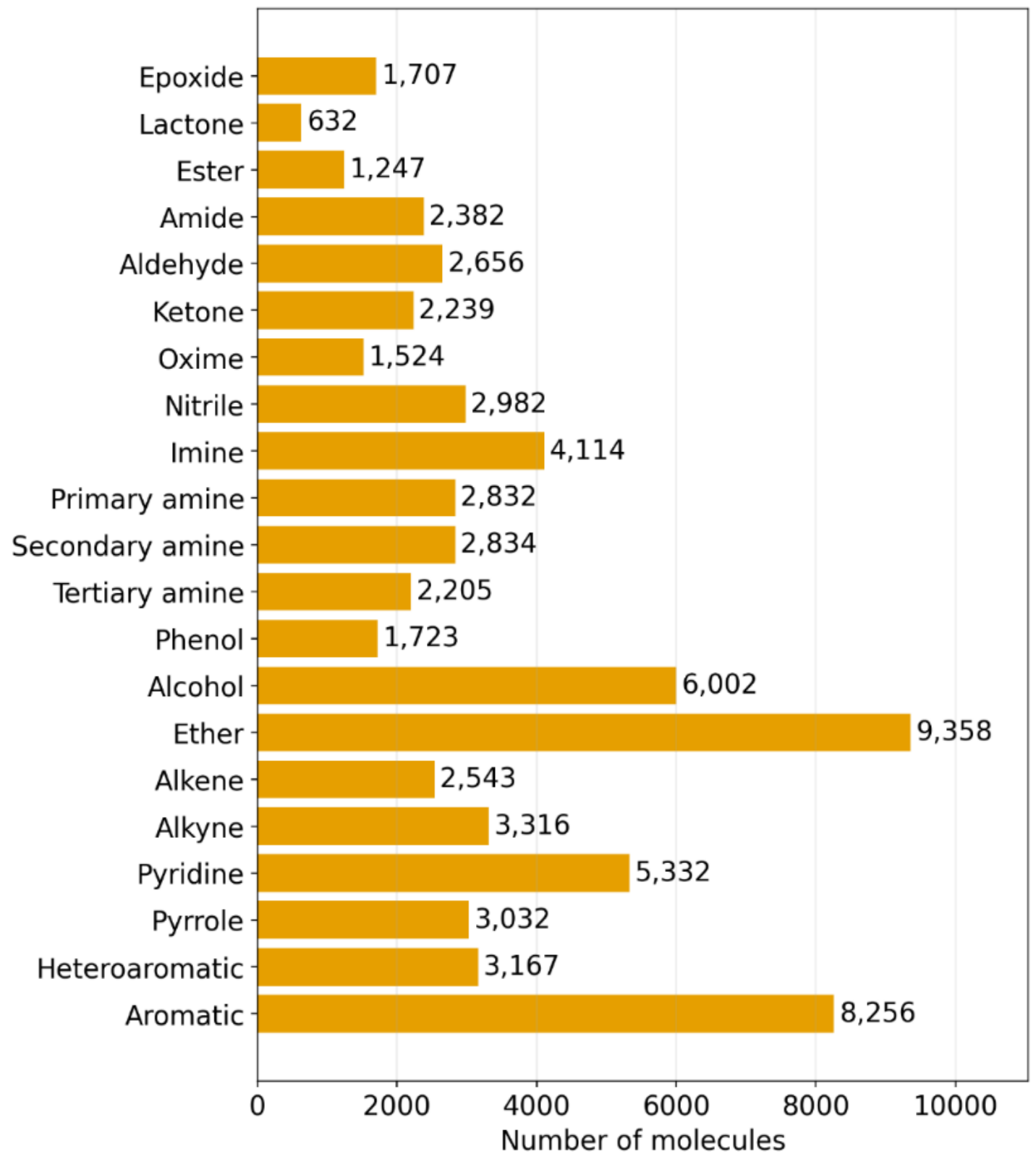

**Figure S1.** Occurrence of functional groups and structural motifs in the molecular dataset. Number of molecules containing each of the 21 functional groups and structural motifs evaluated in Figure 2c. Because a molecule may contain multiple functional groups or structural motifs, the categories are not mutually exclusive, and the counts may overlap.

## Site-Specific XANES-Conditioned Generation of Local Molecular Structures

### Methods

In addition to whole-molecule XANES conditioning, we performed molecular structure generation using site-specific XANES spectra associated with individual absorbing atoms. The model architecture and training procedure followed those of the inverse model described in the main text, with modifications to the conditioning scheme and loss function to explicitly associate each input spectrum with its absorbing site.

For site-specific generation, a single C, N, or O atom was designated as the absorbing site, and its corresponding XANES spectrum was used as the conditioning input. The atomic indices were reordered to place the absorbing atom at index 0. Spectral features extracted by the encoder were provided to the molecular generator and explicitly incorporated into the node representation of the designated absorbing atom. During generation, the total atom count, including hydrogen atoms, and the elemental identity of the absorbing atom were specified, whereas the atomic species of the remaining atoms, bond connectivity, and 3D atomic coordinates were generated by the model.

To improve local structural reconstruction, an additional local loss was introduced alongside the standard reconstruction loss. This loss was evaluated for atoms and bonds within two bonds of the absorbing atom in the reference molecular graph and comprised reconstruction errors for atomic coordinates, atomic species and formal charges, and bond information. The additional loss emphasized the local environments associated with site-specific XANES while retaining the objective of whole-molecule structure generation. Ten molecular candidates were generated for each absorbing site.

Local structural recovery was evaluated using elemental-composition matching, local root-mean-square deviation (RMSD), and functional-group F1 scores. For elemental-composition matching, the first through fourth coordination shells were defined by the shortest bond-path distance from the absorbing atom in the molecular graph. At each shell radius, the elemental identities and counts of atoms within the corresponding range were compared with those of the reference structure. A match was recorded if at least one of the ten candidates reproduced the reference elemental composition.

Local RMSD was evaluated for candidates with matching elemental compositions within the corresponding coordination shells. The absorbing atom was fixed at the origin, and atomic correspondences between identical elements and rotational alignment were optimized using the Kabsch algorithm[1]. The absorbing atom itself was excluded from the RMSD calculation, and the lowest RMSD among the ten candidates was used for each input.

Functional-group recovery was assessed based on the local bonding environment of the designated absorbing atom, rather than the whole-molecule multilabel classification used in the main text. Following bond reconstruction, functional groups were identified using SMARTS patterns, and groups containing the absorbing atom were classified according to a predefined priority order. The predicted labels obtained from the ten generated candidates were compared with the reference label for each absorbing site to calculate category-wise F1 scores and their macro-average.

**Results and Discussion**

The results of site-specific molecular structure generation conditioned on C, N, and O K-edge XANES are presented in Figures S2, S3, and S4, respectively. All three absorption edges showed good recovery of elemental composition, 3D atomic arrangements, and functional groups in the vicinity of the absorbing atom.

We first evaluated elemental-composition matching within the first through fourth coordination shells (Figures S2a, S3a, and S4a). High matching rates were obtained for the first coordination shell across all three absorption edges, followed by a gradual decrease as the evaluated region expanded. This trend is consistent with the greater sensitivity of XANES to the electronic and bonding environments near the absorbing atom. The increasing number of atoms included in the matching criterion at larger shell radii may also contribute to the lower matching rates. Nevertheless, reference-matching elemental compositions were obtained for many molecules even within the fourth coordination shell, indicating that site-specific XANES can provide useful constraints on atomic composition beyond the immediate coordination environment.

We next assessed the local RMSD of candidates with matching elemental compositions (Figures S2b, S3b, and S4b). For all absorption edges, many candidates exhibited small RMSDs within the first coordination shell, whereas the distributions broadened toward larger values as more distant shells were included. This behavior suggests that site-specific XANES provides stronger geometric constraints near the absorbing atom, while more distant regions can adopt different configurations while preserving similar local bonding environments. Representative structural comparisons (Figures S2c, S3c, and S4c) further illustrate the accurate reconstruction of atomic arrangements around the absorbing sites, confirming that site-specific spectral information can guide the generation of local 3D molecular geometries.

Finally, functional-group recovery was evaluated for the designated absorbing sites (Figures S2d, S3d, and S4d). High F1 scores were obtained across diverse local bonding environments for all three absorption edges, indicating that the generated structures reproduced not only the elemental composition and geometry but also characteristic bonding configurations associated with the absorbing elements. Collectively, these results demonstrate that site-specific XANES conditioning enables consistent reconstruction of local molecular structures across different absorbing elements, supporting the use of element-specific spectroscopic information for 3D molecular structure generation.

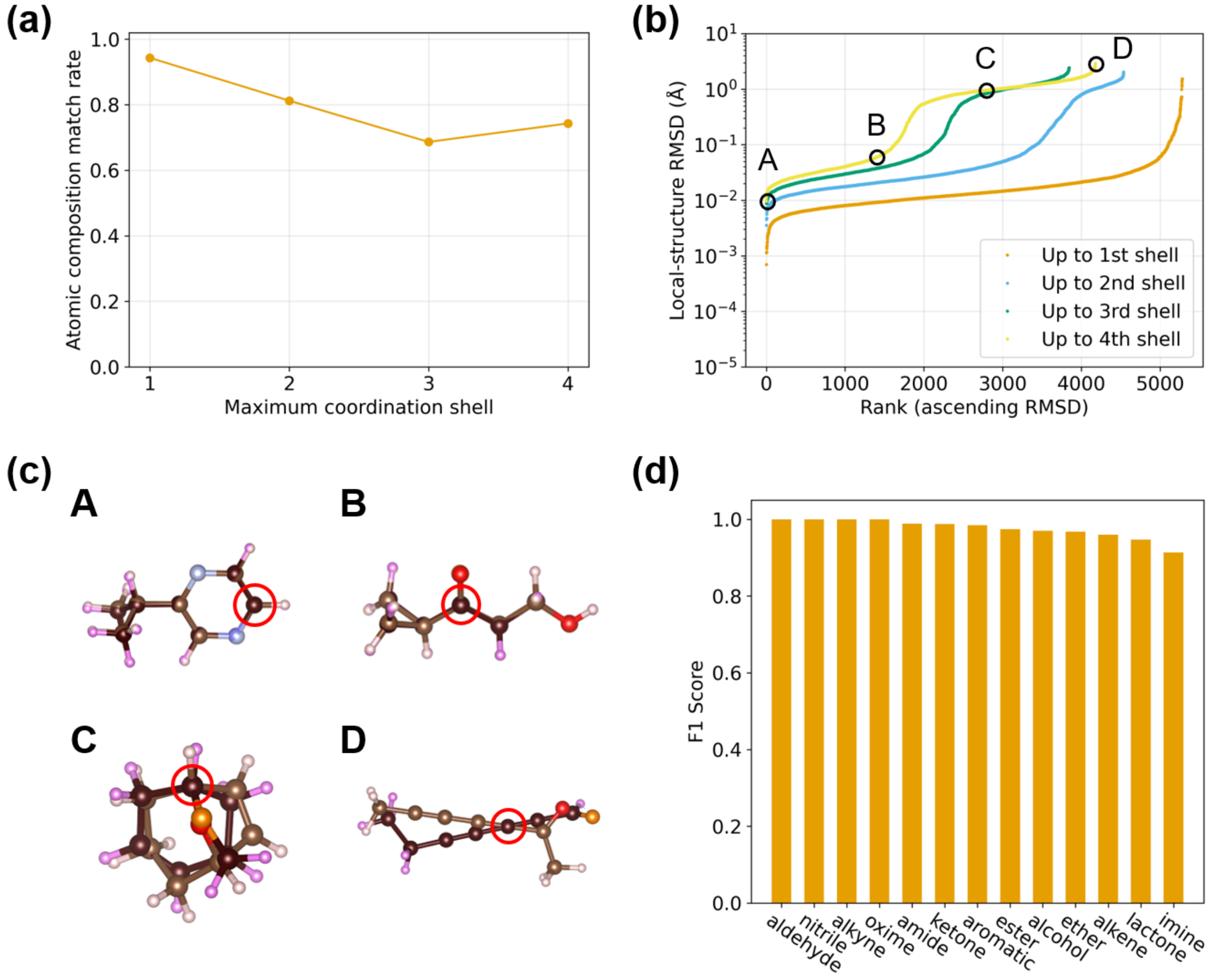

**Figure S2.** Local molecular structure generation from site-specific C K-edge XANES. (a) Best-of-10 match rates of elemental compositions within the first to fourth coordination shells around the absorbing atom. (b) Distributions of best-of-10 local-structure root-mean-square deviations (RMSDs) after alignment centered on the absorbing atom, sorted independently in ascending order for each coordination shell. (c) Comparisons of reference and generated molecular structures at the positions marked A–D in b. In the reference structures, H, C, N, and O atoms are shown in light pink, brown, light blue, and red, respectively. In the generated structures, the corresponding atoms are shown in dark pink, dark brown, blue, and orange, respectively. Red circles indicate the absorbing atoms. (d) F1 scores for carbon-containing functional groups associated with the absorbing sites.

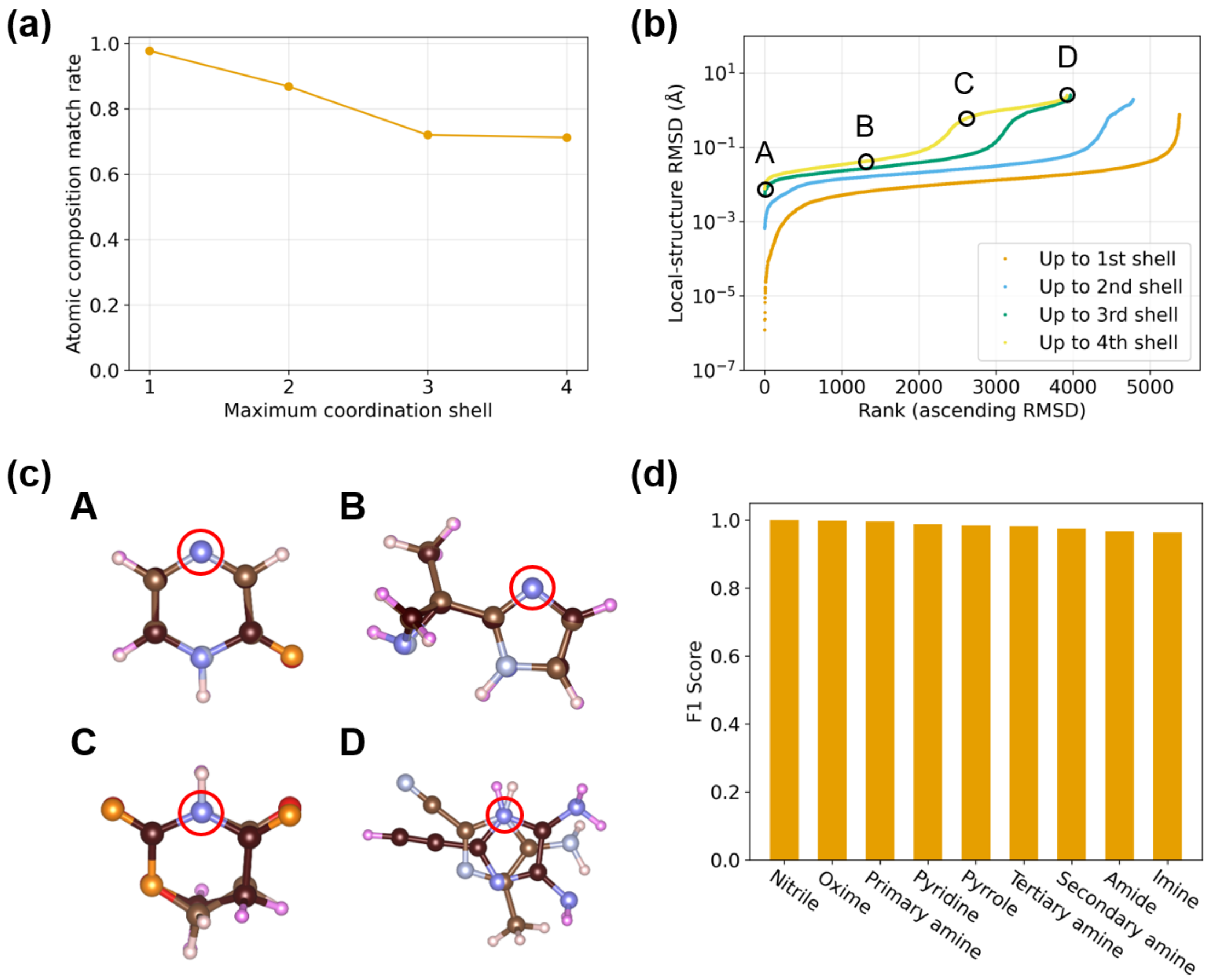

**Figure S3.** Local molecular structure generation from site-specific N K-edge XANES. (a) Best-of-10 match rates of elemental compositions within the first to fourth coordination shells around the absorbing atom. (b) Distributions of best-of-10 local-structure root-mean-square deviations (RMSDs) after alignment centered on the absorbing atom, sorted independently in ascending order for each coordination shell. (c) Comparisons of reference and generated molecular structures at the positions marked A–D in b. In the reference structures, H, C, N, and O atoms are shown in light pink, brown, light blue, and red, respectively. In the generated structures, the corresponding atoms are shown in dark pink, dark brown, blue, and orange, respectively. Red circles indicate the absorbing atoms. (d) F1 scores for nitrogen-containing functional groups associated with the absorbing sites.

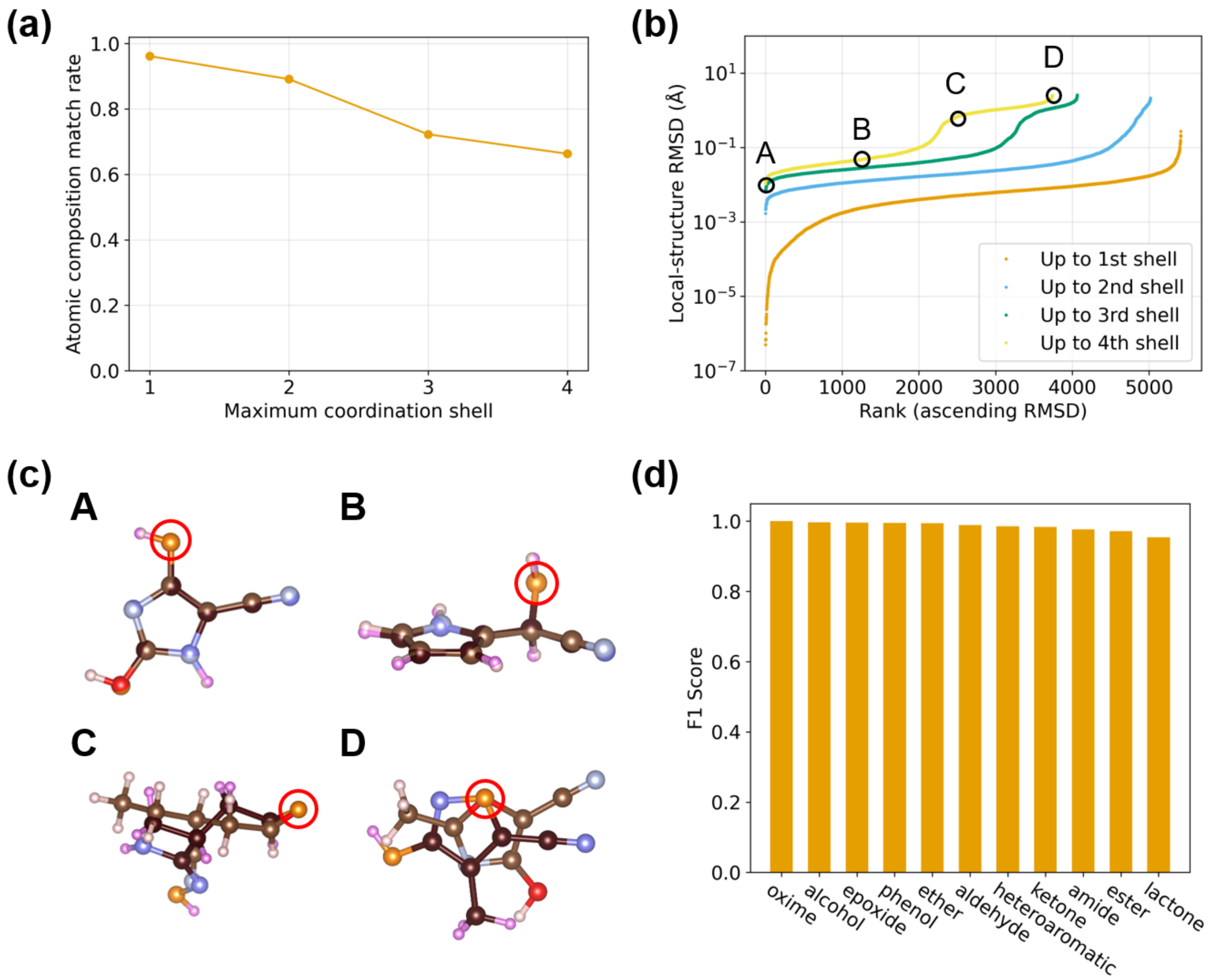

**Figure S4.** Local molecular structure generation from site-specific O K-edge XANES. (a) Best-of-10 match rates of elemental compositions within the first to fourth coordination shells around the absorbing atom. (b) Distributions of best-of-10 local-structure root-mean-square deviations (RMSDs) after alignment centered on the absorbing atom, sorted independently in ascending order for each coordination shell. (c) Comparisons of reference and generated molecular structures at the positions marked A–D in b. In the reference structures, H, C, N, and O atoms are shown in light pink, brown, light blue, and red, respectively. In the generated structures, the corresponding atoms are shown in dark pink, dark brown, blue, and orange, respectively. Red circles indicate the absorbing atoms. (d) F1 scores for oxygen-containing functional groups associated with the absorbing sites.

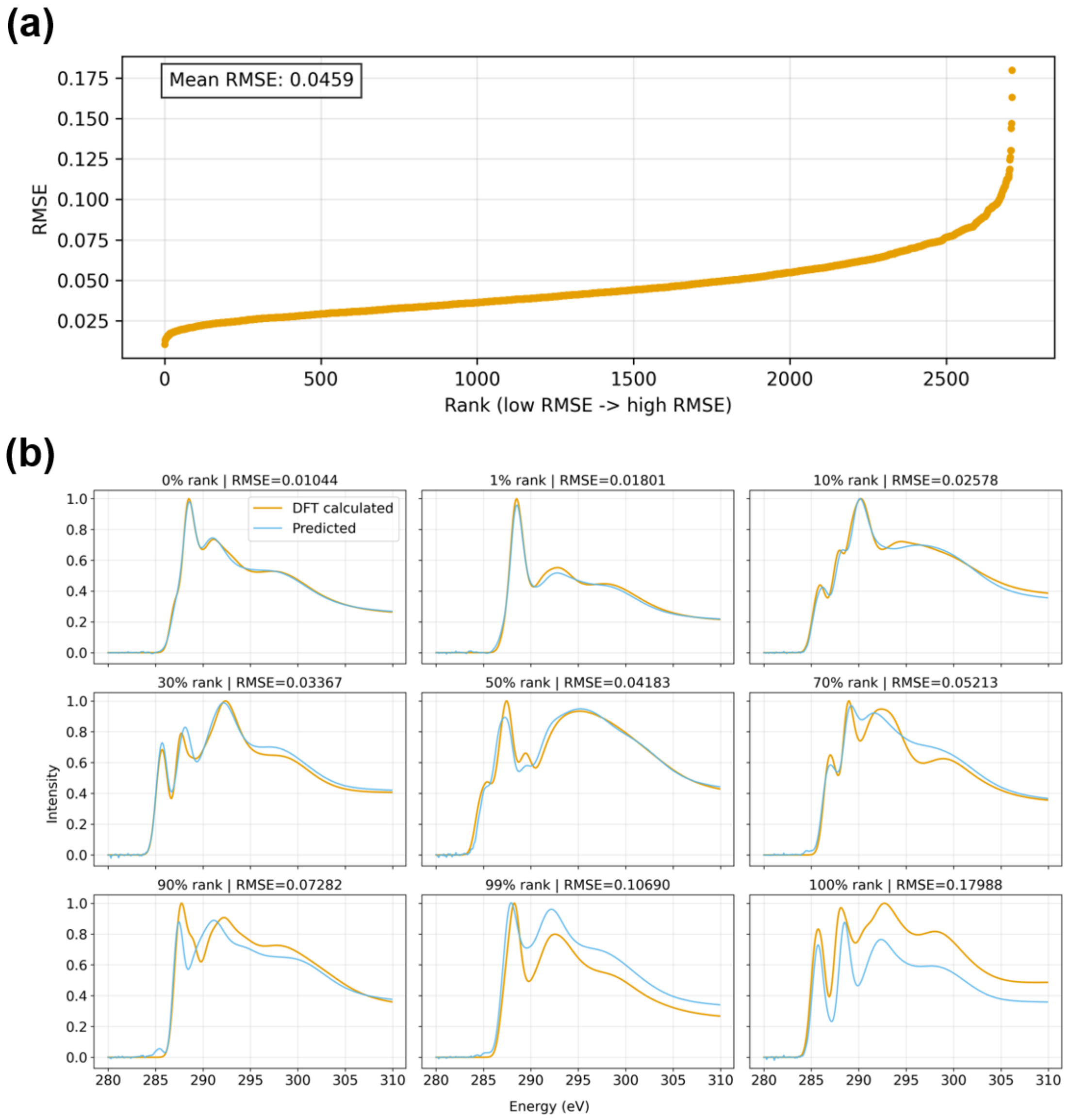

**Figure S5.** Accuracy of the forward XANES model for the C K-edge. (a) Distribution of root-mean-square errors (RMSEs) for test molecules, sorted in ascending order (mean RMSE = 0.0459). (b) Comparison of DFT-calculated (orange) and forward-model-predicted (blue) XANES spectra for representative molecules selected at the 0%, 1%, 10%, 30%, 50%, 70%, 90%, 99%, and 100% positions of the ranked RMSE distribution.

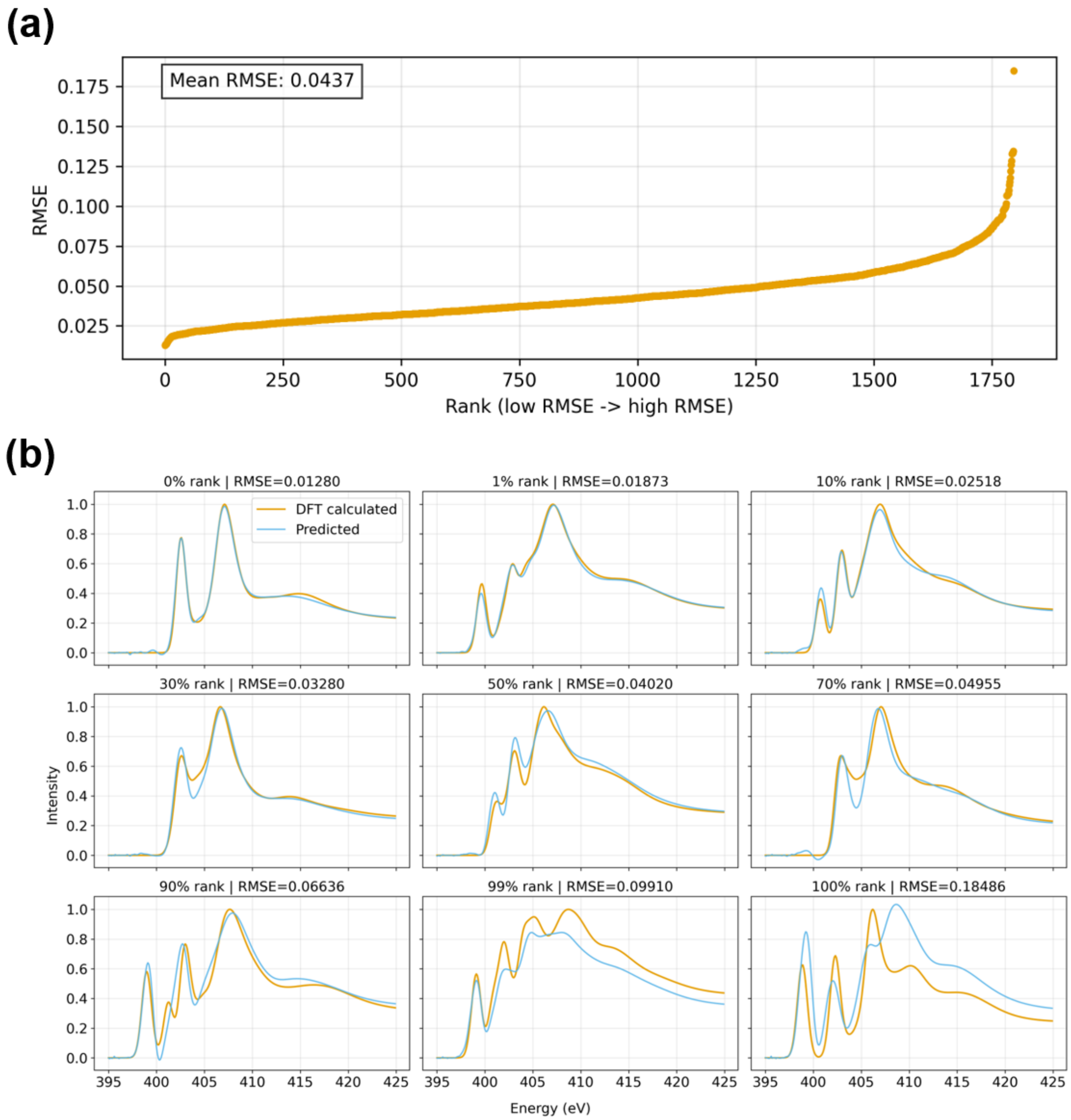

**Figure S6.** Accuracy of the forward XANES model for the N K-edge. (a) Distribution of root-mean-square errors (RMSEs) for test molecules, sorted in ascending order (mean RMSE = 0.0437). (b) Comparison of DFT-calculated (orange) and forward-model-predicted (blue) XANES spectra for representative molecules selected at the 0%, 1%, 10%, 30%, 50%, 70%, 90%, 99%, and 100% positions of the ranked RMSE distribution.

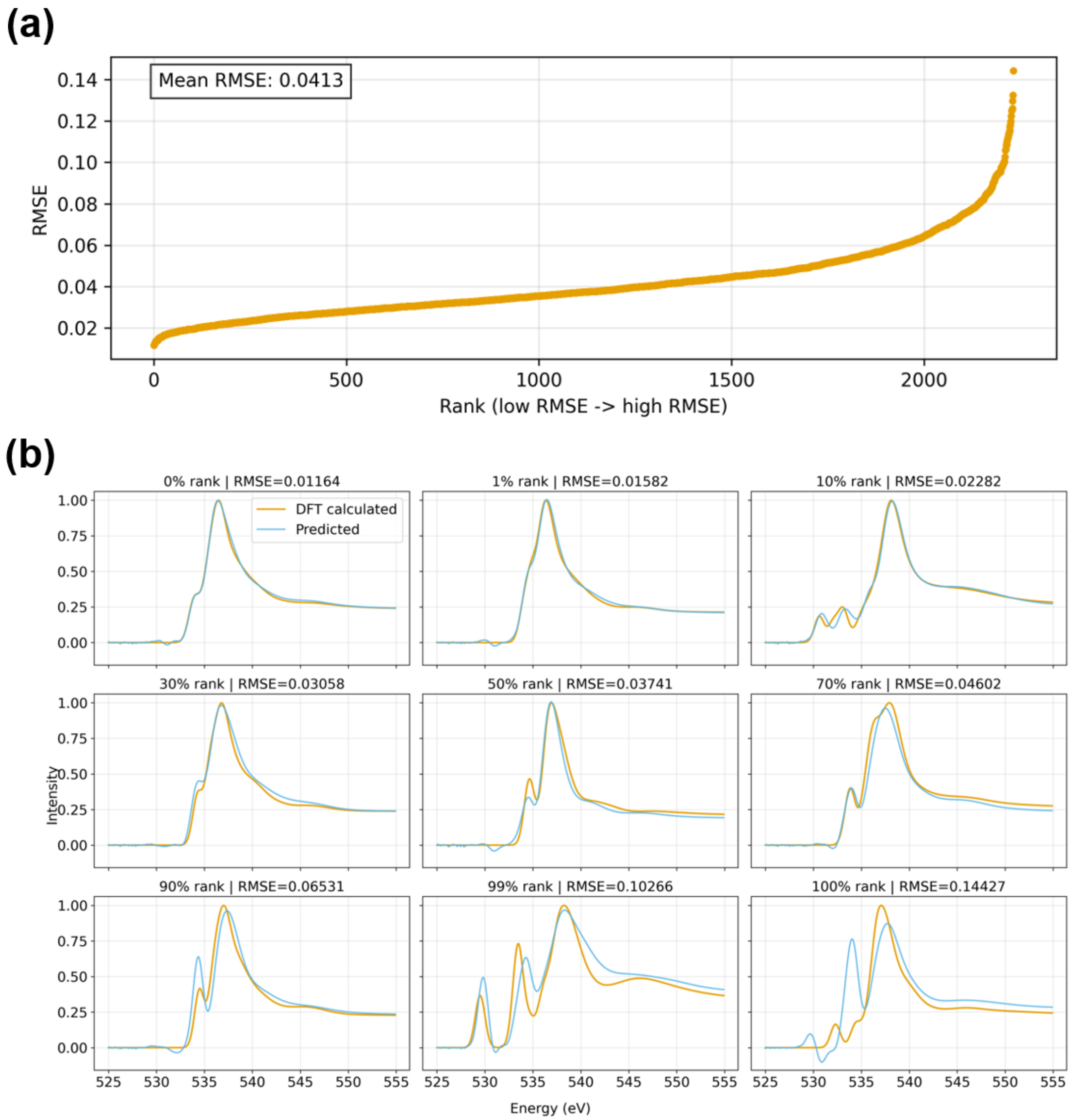

**Figure S7.** Accuracy of the forward XANES model for the O K-edge. (a) Distribution of root-mean-square errors (RMSEs) for test molecules, sorted in ascending order (mean RMSE = 0.0413). (b) Comparison of DFT-calculated (orange) and forward-model-predicted (blue) XANES spectra for representative molecules selected at the 0%, 1%, 10%, 30%, 50%, 70%, 90%, 99%, and 100% positions of the ranked RMSE distribution.

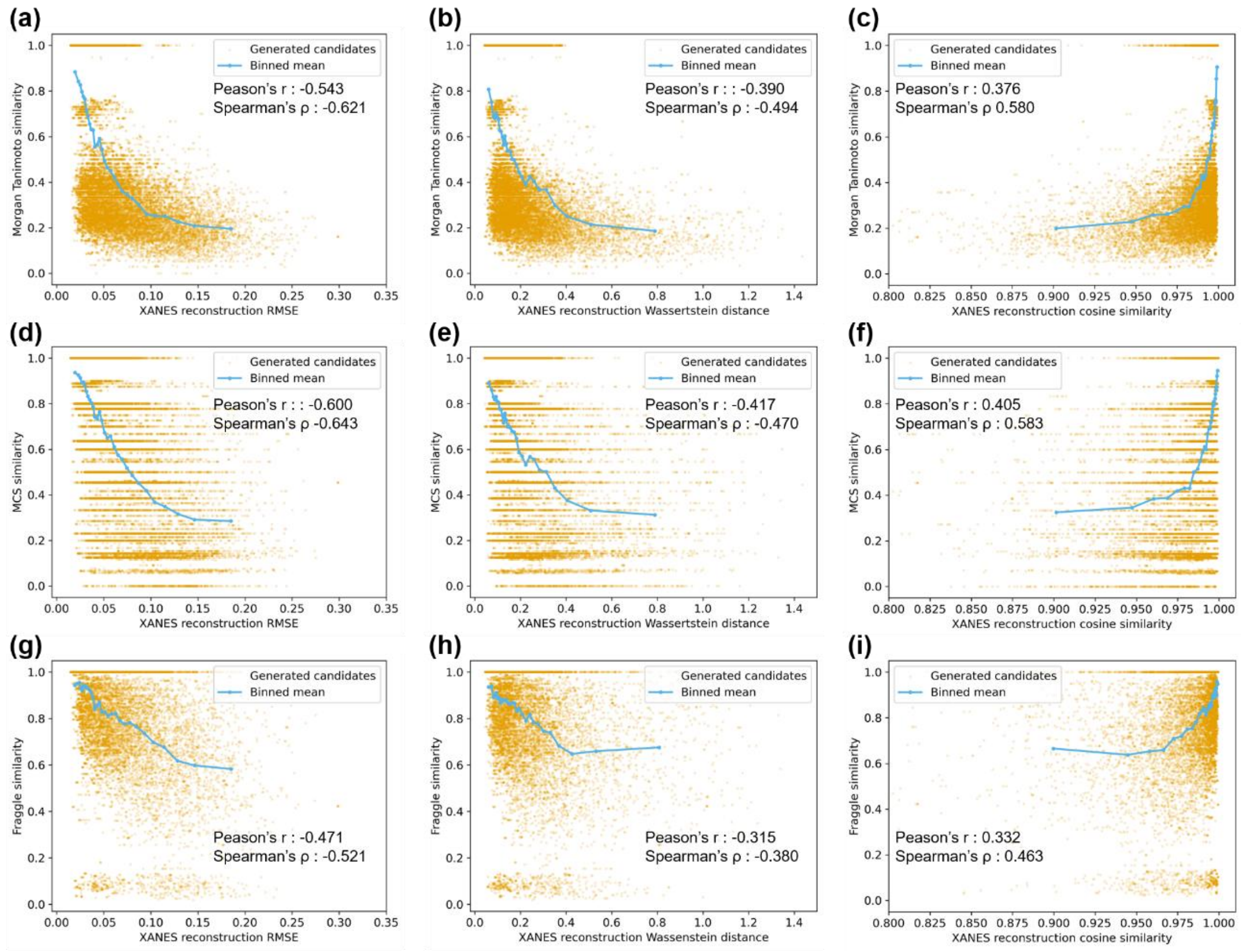

**Figure S8.** Correlation between spectral reconstruction metrics and molecular similarity. Relationships between molecular structural similarities (Morgan Tanimoto, MCS, and Fraggle) and XANES spectral similarity metrics (RMSE, Wasserstein distance, and cosine similarity). Rows correspond to the three molecular similarity metrics, while columns represent the three spectral metrics. Orange points represent individual generated candidates, and blue lines indicate the mean molecular similarity within quantile-based bins of each spectral metric. Pearson correlation coefficients ($r$) and Spearman rank correlation coefficients ($\rho$) are shown in each panel.